\documentclass[aps,prb,twocolumn,floatfix,superscriptaddress]{revtex4-1}

\usepackage[english]{babel}
\usepackage[utf8x]{inputenc}
\usepackage{amsmath}
\usepackage{graphicx}
\usepackage{dcolumn}
\usepackage[caption=false]{subfig}
\usepackage[colorinlistoftodos]{todonotes}

\begin{document}
\title{Spin-Hall Torques Generated by Rare-Earth (Lanthanide) Thin Films}
\author{Neal Reynolds}
\affiliation{Cornell University, Ithaca, New York 14853, USA}
\author{Priyamvada Jadaun}
\affiliation{Cornell University, Ithaca, New York 14853, USA}
\author{John T. Heron}
\affiliation{University of Michigan, Ann Arbor, Michigan 48109, USA}
\author{Colin L. Jermain}
\affiliation{Cornell University, Ithaca, New York 14853, USA}
\author{Jonathan Gibbons}
\affiliation{Cornell University, Ithaca, New York 14853, USA}
\author{Robyn Collette}
\affiliation{Cornell University, Ithaca, New York 14853, USA}
\author{R. A. Buhrman}
\affiliation{Cornell University, Ithaca, New York 14853, USA}
\author{D. G. Schlom}
\author{D. C. Ralph}
\affiliation{Cornell University, Ithaca, New York 14853, USA}
\affiliation{Kavli Institute at Cornell for Nanoscale Science, Ithaca, NY 14853, USA}

\begin{abstract}
We report an initial experimental survey of spin-Hall torques generated by the rare-earth metals Gd, Dy, Ho, and Lu, along with comparisons to first-principles calculations of their spin Hall conductivities. Using spin torque ferromagnetic resonance (ST-FMR) measurements and DC-biased ST-FMR, we estimate lower bounds for the spin-Hall torque ratio, $\xi_{SH}$, of $\approx$ 0.04 for Gd, $\approx$ 0.05 for Dy, $\approx$ 0.14 for Ho, and $\approx$ 0.014 for Lu. The variations among these elements are qualitatively consistent with results from first principles (density functional theory, DFT, in the local density approximation with a Hubbard-U correction). The DFT calculations indicate that the spin Hall conductivity is enhanced by the presence of the partially-filled $f$ orbitals in Dy and Ho, which suggests a strategy to further strengthen the contribution of the $f$ orbitals to the spin Hall effect by shifting the electron chemical potential.   
\end{abstract}

\maketitle

\section{Introduction}

In materials with strong spin-orbit coupling, the spin Hall effect (SHE) allows an applied charge current to generate a transverse spin current \cite{DYAKONOV1971459,PhysRevLett.83.1834,PhysRevLett.85.393,PhysRevLett.92.126603}.  The SHE can be used to apply efficient spin-transfer torques in heavy-metal/ferromagnet bilayers for use in controlling magnetic memory devices \cite{Liu555,PhysRevLett.106.036601,natmat12emori,natnano8ryu}. The relative efficiency of different materials in generating spin-Hall torques is characterized by a figure of merit we will call the spin-Hall torque ratio: $\xi_{SH} = T_{int} \theta_{SH} = T_{int} (\hbar/2e) J_S/J_e$, where $T_{int}$ is a spin transparency factor for the heavy-metal/ferromagnet interface, $\theta_{SH}$ is the spin Hall ratio describing the efficiency of spin-current generation within the heavy metal, $J_S$ is the spin current density generated within the heavy metal, and $J_e$ is the applied charge current density \cite{PhysRevB.92.064426}. Since in general $T_{int} < 1$, we expect $\xi_{SH} < \theta_{SH}$. Values for $\xi_{SH}$ and/or $\theta_{SH}$ have by now been measured for many materials. In pure elements, the strongest spin Hall effects observed \cite{PhysRevLett.112.197201,PhysRevB.90.140407,apl.104.20.10.1063} have been found in the $5d$ transition metals, specifically Pt (with $\xi^{Pt}_{SH} = 0.06$ for Pt/Permalloy samples \cite{PhysRevLett.106.036601} and $0.15$ for Pt/FeCoB) \cite{PhysRevB.93.144409,PhysRevB.92.064426}, $\beta$-Ta ($\xi^{\beta-Ta}_{SH} = -0.12$) \cite{Liu555}, and $\beta$-W ($\xi^{\beta-W}_{SH} = -0.30$) \cite{apl.101.12.10.1063}. The SHE in these pure metals is believed to be largely intrinsic in nature, arising from the Berry curvature of the metal's band structure \cite{PhysRevLett.92.126603,PhysRevB.77.165117,PhysRevB.80.153105}. Work on alloys has shown that extrinsic scattering from defects can also produce a large SHE \cite{PhysRevLett.104.186403,PhysRevLett.106.157208,PhysRevB.89.054401,PhysRevB.2.4559,PhysRevB.64.014416,physica.24,ProgTheorPhys.128.805}. For example, for  Cu(Bi) \cite{PhysRevLett.109.156602} $\theta^{Cu(Bi)}_{SH} = -0.24$ and for Cu(Ir) \cite{PhysRevLett.106.126601} $\theta^{Cu(Ir)}_{SH} = 0.021$ \cite{RevModPhys.87.1213}.

\begin{figure}[b]
\includegraphics{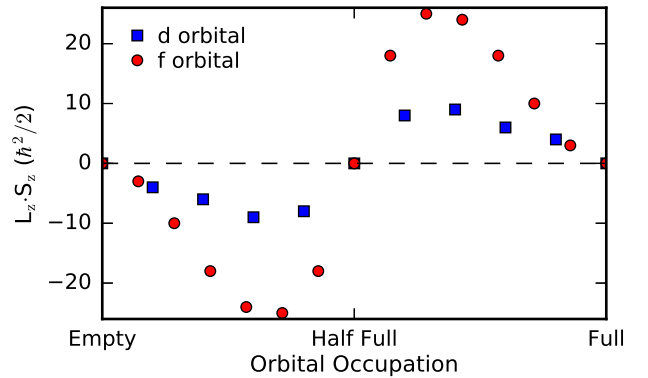}
\caption{\label{fig:LS}(Color online) A naive estimate of the spin-orbit coupling based on Russel-Saunders or `LS' coupling for estimates of the orbital and spin angular momentum components of the total angular momentum for Hund's rules for d and f orbital fillings. }
\end{figure}

Here we investigate the strength of the spin-Hall torque generated by a class of elements that has been relatively unexplored, the $f$-electron lanthanide series or ``rare-earth'' (RE) metals.  Our work is motivated by theoretical predictions that the spin Hall effect in metals with partially full $f$ ($l=3$) orbitals is potentially quite large \cite{PhysRevB.81.224401}.  In the most naive consideration of the SHE, one might expect that the combination of both large orbital angular momenta and large spins in $f$-electron atoms might lead to large spin-orbit coupling terms, scaling approximately as $\langle l \cdot s \rangle$, and consistent with the qualitative trend observed in the 3d, 4d, and 5d transition metals \cite{PhysRevLett.102.016601,:/content/aip/journal/jap/117/17/10.1063/1.4913813,PhysRevB.90.140407}. Within this simple ansatz, rare-earths near $1/4$ and $3/4$ filling of the $f$ orbitals are likely candidates for a strong SHE (Fig.~\ref{fig:LS}).  We caution, however, that the details of the material band structure and its associated Berry curvature can be critical, as well, so that one should not expect simply that $\xi_{SH} \propto \langle l \cdot s \rangle$.  

We report both experimental measurements of spin-Hall torque ratios and theoretical calculations within density functional theory (DFT) of the spin Hall effect in four rare-earth metals: Gd, Dy, Ho, and Lu which have the $f$-level configurations f$^7$ (half-full), f$^9$, f$^{10}$, and f$^{14}$ (full), respectively\cite{PhysRevB.20.1315,Nature399-1999}. These metals were chosen due to the range of orbital fillings that they cover, and because of their relative chemical stability and low vapor pressure (relative to the early lanthanides) \cite{meurant1980vacuum}. To ensure that our measurements are robust, we report spin torque ratios derived from two separate types of measurements: (i) based on the amplitude of spin-torque ferromagnetic resonance (ST-FMR) peaks \cite{PhysRevLett.106.036601,Liu555,Nature.511.449} and (ii) based on the change in the ST-FMR linewidth as a function of an applied DC current \cite{PhysRevLett.106.036601, Nature.511.449,PhysRevLett.101.036601}.  Our results for Gd, Dy, and Ho should be considered  lower bounds on $\xi_{SH}$ because the samples for these materials contained a thin Hf spacer layer between the rare-earth and the ferromagnet to reduce the magnetic damping enough for quantitative ST-FMR experiments.  This spacer layer likely reduces the transparency factor $T_{int}$ for spin transmission from the rare-earth to the ferromagnet.

What we find experimentally is that the spin Hall effect is small, as expected, in half-filled ($\xi^{Gd}_{SH} \approx 0.04 \pm 0.01$) and fully-filled ($\xi^{Lu}_{SH} \approx 0.014 \pm 0.002$) $f$-level configurations, while for the materials with partially-filled $f$-levels the spin torque ratio can be enhanced, $\xi^{Dy}_{SH} \approx 0.05 \pm 0.01$ and $\xi^{Ho}_{SH} \approx 0.14 \pm 0.02$. The magnitude of the value we find for Ho is among the largest for any element (comparable to Pt and Ta and smaller only than $\beta$-W \cite{apl.101.12.10.1063}). The sign we measure for $\xi_{SH}$ is positive in all four cases (the same as Pt), as expected based on Hund's rules given the orbital filling. Our results for Gd differ from a recent report by the Beach group in both sign and magnitude \cite{apl.108.232405}.  

Our first-principles calculations suggest that the degree to which $f$ orbitals affect the spin-Hall torque ratio depends on the proximity of the  $f$ levels to the Fermi level, $\epsilon_F$.  In Lu the $f$ orbitals are well below $\epsilon_F$ such that the spin Hall effect is determined entirely by the contributions from the $d$ orbitals.  In Dy and Ho, however, the $f$ levels are sufficiently close to $\epsilon_F$ as to provide an enhanced spin Hall effect.  
Our calculations suggest that further $f$-orbital enhancement of $\xi_{SH}$ may be possible by artifically shifting the $f$-electron density of states closer to $\epsilon_F$, e.g., by alloying. 

\section{Experiment}

\subsection{Materials Growth and Device Fabrication}

For our Lu devices, we used molecular-beam deposition (MBD) to grow multilayer samples consisting of substrate/Fe (5 nm)/Lu (10 nm)/Al (2.5 nm) using a sapphire(0001) substrate. The entire structure was deposited without heating the substrate, \textit{i.e.} at room temperature. The Al protective cap was then oxidized on exposure to air. Fe was chosen as the magnetic layer because of its availability in the MBD system. This stack ordering is inverted compared to most studies of the SHE in heavy-metal/ferromagnet bilayers, because we found that it was necessary to grow the Fe layer first to produce smooth, continuous films. Attempts to to grow Lu by MBD directly on a sapphire substrate resulted in substantial islanding.  For the other rare-earths, we were not able to grow any successful samples by MBD. Growth of Ho directly on sapphire yielded even worse island formation than Lu, and when we attempted to study substrate/Fe/Ho/Al samples grown by MBD the Al did not provide enough protection to keep the Ho sample from oxidizing. This could be directly observed as a progressive color change  that propagated inward from the corners of the 10 mm x 10 mm chip to its center over the course of $\approx$15 seconds as soon as the sample was removed from the loadlock of the MBD system. 

To surmount this problem, for our Gd, Dy, and Ho samples we switched to DC magnetron sputtering using a high-vacuum sputter system (base pressure $<$5 $\times$ $10^{-9}$ Torr). The rare-earth layers were sputtered from a 1'' target positioned directly beneath the center of the sample plate. Depositions were conducted without heating the substrate, \textit{i.e.} at room temperature. Sputtering of the RE metals onto a sapphire substrate at room temperature yielded smooth films with RMS roughness $<$ 500 pm by atomic force microscopy. We were then able to grow smooth, continuous films of Permalloy (Py $=$ Ni$_{80}$Fe$_{20}$) on top of the RE followed by a 2 nm Al cap. Unfortunately, when we attempted ferromagnetic resonance measurements on the Dy and Ho devices no resonance could be detected, presumably because the damping was very large.  Previous studies of Py films containing RE impurities have found a similar trend in the magnitude of damping as a function of varying the RE element --  RE elements for which the orbital angular momentum contribution to the RE moment is non-zero (like Dy and Ho, but not Gd) can greatly increase spin relaxation to the lattice \cite{IEEEtransmagn.37.2001,apl/82/8/10.1063/1.1544642}. 

To counteract this increased damping, we grew samples with a Hf spacer layer inserted between the RE=Gd,Dy,Ho layer and the Py layer to minimize intermixing and reduce the magnetic damping \cite{apl.hfenhances}. While not necessary, the Gd samples also received this Hf spacer layer to enable us to make robust statements about any observed trend in spin Hall effect. Ultimately, for RE=Gd, Dy, and Ho the multilayer stack we used for the measurements was RE (10 nm)/Hf ($t_{Hf}$)/Py (5 nm)/Al (2.5 nm) grown at room temperature by sputtering onto (0001) sapphire, with the Al oxidized in air.  Most of our measurements were performed with $t_{Hf}$ = 1.5 nm, although we did perform some tests with other thicknesses as described below. X-ray diffraction experiments of the sputtered samples indicate that both the RE metal and the Py film are polycrystalline with randomly oriented grains.  Control samples with thicker Hf layers [both Ho (10 nm)/Hf (4 nm)/Py (5 nm)/Al (2.5 nm) and Hf (5 nm)/Py (5 nm)/Al (2.5 nm)] exhibited Hf films with notable (0001) fiber texture.  Atomic force microscopy indicated smooth films with RMS roughness $<$ 500 pm for all of these multilayer samples. 

We used optical lithography and Ar ion milling to define bars 30 $\mu$m long by 8 $\mu$m wide. To finish device fabrication, we made contact pads (Ti (30 nm)/Pt (250 nm)) via optical lithography and liftoff. There was no color change observed in any of the thin film samples upon exposure to air; all remained visibly shiny and metallic throughout fabrication and processing. 

\begin{figure*}
	\subfloat{
        \includegraphics{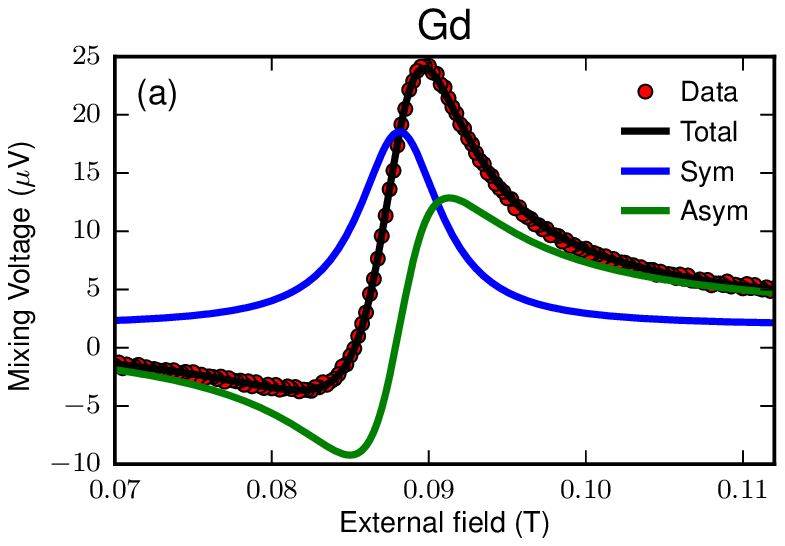}
        \label{fig:Gd_STFMR}
        }   
    \subfloat{
        \includegraphics{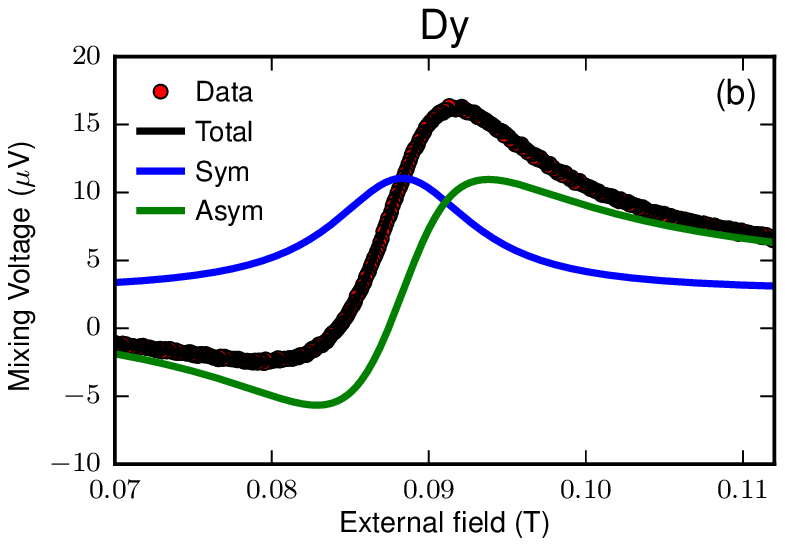}
        \label{fig:Dy_STFMR}
        }
  
    \subfloat{
        \includegraphics{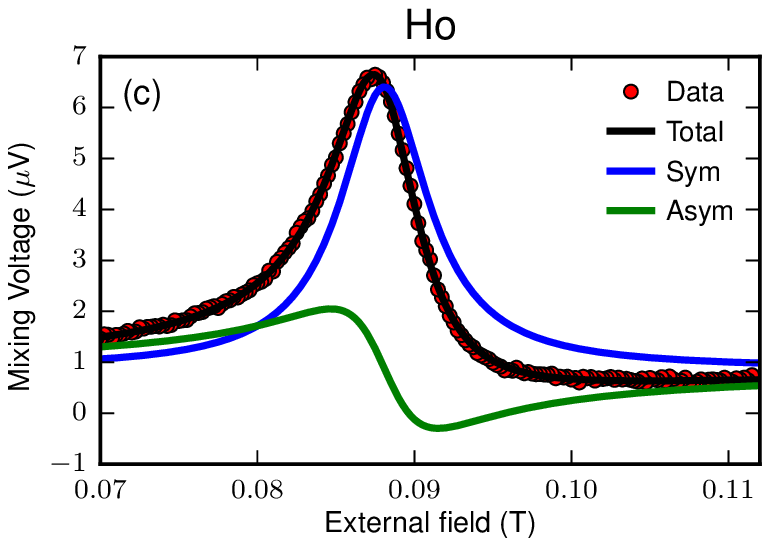}
        \label{fig:Ho_STFMR}
        }     
    \subfloat{
        \includegraphics{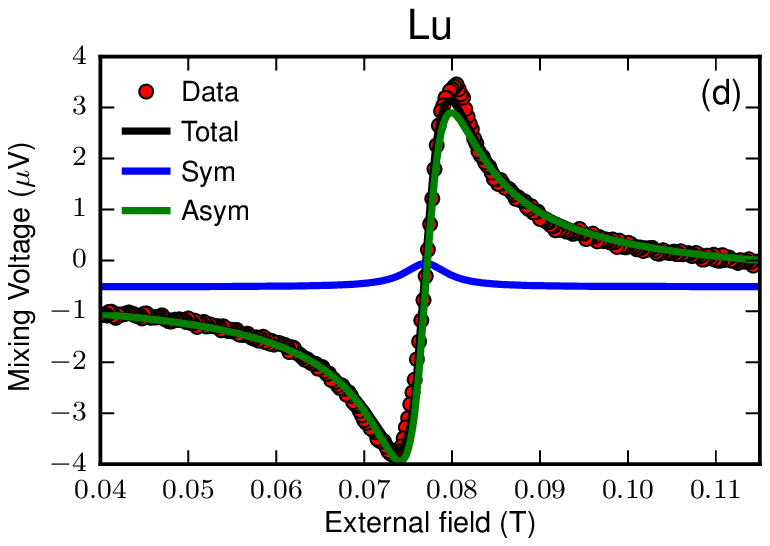}
        \label{fig:Lu_STFMR}
        }
    \caption{(Color online) Representative ST-FMR resonance curves at 9 GHz for (a) a Gd(10)/Hf(1.5)/Py(5)/AlO$_x$ sample (numbers in parentheses are thicknesses in nm), (b) a Dy(10)/Hf(1.5)/Py(5)/AlO$_x$ sample, and (c) a Ho(10)/Hf(1.5)/Py(5)/AlO$_x$ sample. (d) Representative trace for a Fe(5)/Lu(10)/AlO$_x$ sample at 11 GHz. The higher frequency used for Lu is due to the larger effective magnetization of the Fe layer in the Fe/Lu as compared to the Py used for all other samples. The sign of the signals generated by the Lu sample can be compared directly to the others despite the ``inverted'' order of the Lu sample (ferromagnet on the bottom, which would ordinarily cause a sign change), because the sign of the anisotropic magnetoresistance in Fe is also reversed relative to the Py in the other samples. The consistent signs of the symmetric components indicate that all four rare-earth systems exhibit a positive spin Hall effect (the same sign as Pt). The sign of the antisymmetric component of the resonance is reversed in the Ho sample relative to the others.}\label{fig:STFMR}
\end{figure*}

\subsection{Measurement}

We performed ST-FMR measurements \cite{PhysRevLett.106.036601,Liu555,Nature.511.449} using a microwave source power of 10 dBm and a fixed frequency in the range 6-12 GHz.  The torques generated by the oscillating microwave-frequency current cause the magnetization orientation $\hat{m}$ of the sample to precess, thereby producing microwave-frequency resistance oscillations on account of the anisotropic magnetoresistance (AMR) of the ferromagnetic layer (Py or Fe). We measured the DC voltage that results from mixing between this oscillating resistance and the oscillating applied current, while sweeping an applied external magnetic field $\overrightarrow{B}$ in the sample plane at an angle approximately 45$^{\circ}$ from the current direction to scan through the resonance condition.  The DC mixing voltage was detected through a bias tee by using lockin detection with amplitude modulation of the RF signal \cite{Liu555,PhysRevLett.106.036601}.  The microwave power absorbed by the devices, and hence the amplitude of the microwave current in the sample, were calibrated using a vector network analyzer to measure the device S$_{11}$ and cabling S$_{21}$ parameters.  Representative ST-FMR traces for each RE are shown in Fig.~\ref{fig:STFMR}.    

The magnetic dynamics during ST-FMR can be modeled using the Landau-Liftschitz-Gilbert-Slonczewski equation \cite{Slonczewski1996L1}
\begin{equation}
\begin{split}
\dot{\hat{m}}&=- \gamma\hat{m}\times(\overrightarrow{B}-(\hat{m} \cdot \hat{z}) \mu_0M_{eff}\hat{z})+\alpha\hat{m}\times\dot{\hat{m}} \\
           & +\frac{\gamma\hbar}{2eM_s}\frac{XI_{RF}}{t_{mag}t_{RE}w}(\xi_{SH}(\hat{m}\times\hat{\sigma}\times\hat{m})-\xi_{\perp}(\hat{m}\times\hat{B}_{Oe}))\label{eqn:LLGS}
\end{split}
\end{equation}
where $\gamma$ is the absolute value of the gyromagnetic ratio, $M_{eff}$ is the effective magnetization which characterizes the out of plane demagnetization field, $\alpha$ is the Gilbert damping parameter, $\hbar$ is the reduced Planck's constant, $e$ is the magnitude of the charge of the electron, $M_s$ is the saturation magnetization of the ferromagnetic layer, $t_{mag}$ is the thickness of the ferromagnetic layer, $t_{RE}$ is the thickness of the RE layer, $w$ is the width of the device, $X$ is the fraction of current that flows through the RE layer determined by measuring the resistivities of the individual layers, $I_{RF}$ is the RF current through the device as calibrated using a vector network analyzer, $\hat{\sigma}$ is spin-moment orientation of the spin current generated by the spin Hall effect, $\hat{B}_{Oe}$ is the direction of the Oersted field for positive current, $\xi_{SH}$ is the in-plane spin torque ratio, and $\xi_{\perp}$ is an out-of-plane (field-like) torque ratio due the Oersted field generated by the RF current and potentially other spin-orbit torques. The oscillating resistance mixes with the applied RF current to give rise to an output voltage signal with components at DC and at twice the applied RF frequency. For an in-plane directed external magnetic field the resulting DC component has the form \cite{Liu555,PhysRevLett.106.036601,Nature.511.449}
\begin{equation}
\begin{split}
V_{mix}=-\frac{I^2_{RF}\gamma^2\hbar l \cos(\phi) X}{4 e M_s Vol_{mag} t_{RE}}&\left(\frac{dR}{d\phi}\right)\times\\
&\left(\xi_{SH}F_S(B)+\xi_{\perp}F_A(B)\right) \label{eqn:fullSTFMRfit}
\end{split}
\end{equation}
\begin{align}
F_S\equiv&\left(\frac{\omega^2\alpha(2B+\mu_0M_{eff})}{(\omega^2-\omega_0^2)^2+\alpha^2\gamma^2\omega^2(2B+\mu_0M_{eff})^2}\right)\\
F_A\equiv&\left(\frac{\gamma^2B(B+\mu_0M_{eff})^2-\omega^2\alpha(B+\mu_0M_{eff})}{(\omega^2-\omega_0^2)^2+\alpha^2\gamma^2\omega^2(2B+\mu_0M_{eff})^2}\right)
\end{align}
where $\phi$ is the angle of the magnetic field relative to the current, $\frac{dR}{d\phi}$ is the amplitude of the resistance oscillation of the device from the AMR of the ferromagnetic layer, and $\omega_0$ is the resonance frequency defined as $\omega_0\equiv\gamma\sqrt{B(B+\mu_0 M_{eff})}$. 

Our first method to determine the spin torque ratios is based on the amplitude of the ST-FMR signal: $\xi_{SH}$ is proportional to the amplitude of the symmetric component of the resonance and $\xi_{\perp}$ is proportional to the antisymmetric part.  As noted above, we calibrate the microwave current in the sample using vector network analyzer measurements, to enable a quantitative determination of the individual torque ratios.  The other materials parameters required to calculate the spin torque ratios from the ST-FMR data are summarized in Table~\ref{table:STFMRparameters}. 

\begin{table}
\caption{\label{table:STFMRparameters}A summary of the materials parameters that enter into Eq.~\ref{eqn:fullSTFMRfit} for each sample.  The saturation magnetization $M_s$ is measured using vibrating sample magnetometry. (The values of $M_s$ for Gd, Dy, and Ho correspond to Py, and the higher value for Lu corresponds to Fe.). }
\begin{tabular}{c l l l c }
\hline
\hline
Element & $\mu_0 M_{s}$ (T) & $I_{rf}$ (mA)& $X$ & $\frac{dR}{d\phi}$ ($\frac{\Omega}{Rad}$) \\
\hline
Gd & 0.88 $\pm$ 0.08	&   6.0 $\pm$ 0.3 &	0.25 &	3.72 	\\
Dy & 0.94 $\pm$ 0.09	&  	6.3 $\pm$ 0.3 &	0.27 &	2.93	\\
Ho & 1.1 $\pm$ 0.1		&	2.8 $\pm$ 0.1 &	0.13 &	3.94	\\
Lu & 2.3 $\pm$ 0.1 		&	9.6 $\pm$ 0.2 &	0.07 &	-0.30	\\
\hline
\hline
\end{tabular}
\end{table}

\begin{figure}[h!]
\centering
\includegraphics{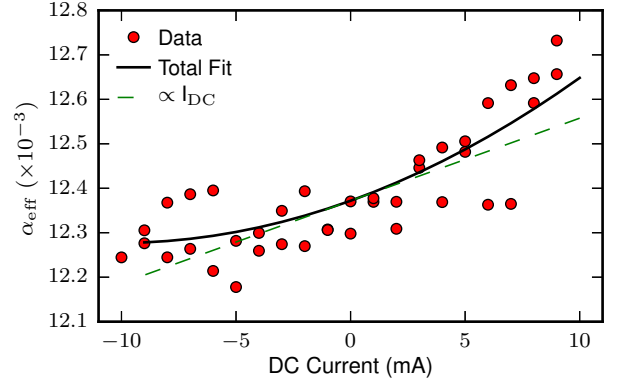}
\caption{\label{fig:Ho_alphas}(Color online) Evolution of damping of a Ho(10)/Hf(1.5)/Py(5)/AlO$_x$ stack at 6 GHz with applied DC current. The linear component yields information about the spin Hall effect, while the quadratic background indicates heating. }
\end{figure}

As a second approach to measure the spin-Hall torque efficiencies, we also performed DC-biased ST-FMR measurements on the same devices for the Gd, Dy, and Ho multilayers.  (This measurement was not successful for the Fe/Lu due to small signal levels associated with much smaller AMR signals in Fe compared to Py.) The presence of a constant DC bias modifies Eq.~\ref{eqn:LLGS} by adding both a DC in-plane (spin Hall) torque and a DC out-of-plane torque. The effect of the in-plane torque is to modify the effective Gilbert damping of the system, increasing or decreasing the linewidth of the ST-FMR resonance depending on the sign of the applied DC current. Assuming that the DC bias is sufficiently small that the damping modification is in the linear regime,  \cite{PhysRevLett.106.036601,Nature.511.449}
\begin{equation}
\xi_{SH}=\frac{d\alpha_{\rm eff}}{d I_{RE}}\frac{\hbar}{2e}\frac{w t_{RE} M_s t_{mag}(B_0+\mu_0M_{\rm eff}/2)}{\sin{\phi}}
\end{equation}
where $\alpha_{\rm eff}$ is the effective damping determined by fitting to the resonance linewidth,  $B_0$ is the resonant field, $w$ is the device width, $t_{mag}$ is the thickness of the magnetic layer, and
 $I_{RE}\equiv X I_{RF}$ is the current through rare-earth layer.

We carried out the DC-biased ST-FMR measurements using field modulation provided by a Helmholtz coil to measure the derivative of the ST-FMR mixing voltage signal. This was necessary to eliminate a large background offset that arises when locking into an amplitude modulated signal due to heating from the RF current.  The field modulation allows for high sensitivity even at large (10 mA) applied currents. A representative trace for the damping as a function of current for a Ho multilayer is shown in Fig.~\ref{fig:Ho_alphas}.  In addition to a linear shift of the damping as a function of DC current, we also observe a quadratic background, which is likely due to Joule heating. For determination of $\xi_{SH}$, we use only the linear contribution to the current dependence, determined by a least-squares fit.

\subsection{Analysis of Experimental Results}

\begin{table}
\caption{\label{table:spinHallsummary} Measured spin torque ratios for Gd(10)/Hf(1.5)/Py(5)/AlO$_x$, Dy(10)/Hf(1.5)/Py(5)/AlO$_x$, Ho(10)/Hf(1.5)/Py(5)/AlO$_x$ and Fe(5)/Lu(10)/AlO$_x$. $\xi_{\perp,SO}$ refers to the spin-orbit-generated part of the field-like torque, with the contribution from the Oersted field subtracted.}
\begin{tabular}{c c c c }
\hline
\hline
Element & Hf thickness (nm) & $\xi_{SH}$ & $\xi_{\perp,SO}$ \\
\hline
&&ST-FMR amplitude &\\
Gd & 1.5 & 0.04$\pm$0.01 & -0.04 \\
Dy & 1.5 & 0.06$\pm$0.01 & -0.03  \\
Ho & 1.5 & 0.16$\pm$0.04  & -0.07 \\
Lu & 0 & 0.014$\pm$0.009  & -0.03 \\
\hline
&&DC-biased ST-FMR&\\
Gd & 1.5 & 0.04$\pm$0.01 & \\
Dy & 1.5 & 0.05$\pm$0.01 &  \\
Ho & 1.5 & 0.12$\pm$0.02 &  \\
\hline
\hline
\end{tabular}
\end{table}

Our results for $\xi_{SH}$ from both the ST-FMR amplitude measurements and the DC-biased ST-FMR measurements are shown in Table ~\ref{table:spinHallsummary} for the Gd, Dy, and Ho samples with a 1.5 nm Hf spacer layer, and for the Lu/Fe samples with no Hf spacer. These data are also plotted at the conclusion of the paper in Fig.~\ref{fig:summaryfigure}a.  We find good agreement between the two types of measurements in the samples for which both measurements could be performed. It is important to note that these results do not correspond to intrinsic values of $\theta_{SH}$ for each rare-earth due to spin current attenuation in the Hf spacer layer and also due to the likelihood of additional spin relaxation at the interfaces.  

For convenience in comparing to theoretical calculations, the strengths of the spin-orbit torques can alternatively be expressed in terms of spin torque efficiencies per unit electric field (or ``spin torque conductivity''), $\sigma^{\rm exp}_{SH} = \xi_{SH} / \rho_{RE}$, where $\rho_{RE}$ is the electrical resistivity of the rare-earth.  We have included these values in Table~\ref{table:spinconductivitysummary} as well as the values of $\rho_{RE}$. For RE=Gd, Dy, and Ho, we determined $\rho_{RE}$ via 4-point resistance measurements on control samples of substrate/RE/AlO$_x$. For RE=Lu, we used control measurements on substrate/Fe/AlO$_x$ to extract $\rho_{Lu}$ from the full substrate/Fe/Lu/AlO$_x$ bilayer by treating the Fe and Lu layers as parallel resistors. These data are plotted at the conclusion of the paper in Fig.~\ref{fig:summaryfigure}b as well. We note that given the factor $\hbar / 2e$ in our definition of $\xi_{SH}$ there is explicitly a factor of 2 in our definition of $\sigma^{\rm exp}_{SH}$ that is not used universally in the literature. Just as for our determination of $\xi_{SH}$, our values of $\sigma^{\rm exp}_{SH}$ will be diminished by any spin attenuation in the Hf spacer or interface spin relaxation, so these values represent a lower bound on the intrinsic spin Hall conductivity within each rare-earth.  

What we find is that the spin-Hall torque ratios for Gd, Dy, and Lu are relatively small.  This is to be expected for Gd and Lu because these materials have $f$-electron occupations f$^7$ (half-full) and f$^{14}$ (full), respectively, with the consequence that the $f$-levels are relatively far from the Fermi level and unlikely to contribute any significant Berry curvature. The spin-Hall torque ratio for Ho (f$^{10}$) is significantly greater.  We find for the Ho samples with the 1.5 nm Hf spacer (using the average value of the two measurement techniques) that $\xi_{SHE}$ = 0.14 $\pm$ 0.03. This is among the largest known values for any pure material, comparable to $\beta$-Ta \cite{Liu555} and Pt (in Pt/CoFeB samples \cite{PhysRevB.93.144409,PhysRevB.92.064426}), and less than only $\beta$-W \cite{apl.101.12.10.1063}.  This large value is despite the likelihood, as we will discuss below, of significant spin relaxation at the interfaces in the RE samples.  The spin-Hall torque conductivity $\sigma^{\rm exp}_{SH}$ of Ho is, however, only slightly enhanced relative to Gd and Dy.  This is a consequence of the very high resistivity of our sputtered Ho films (see Table~\ref{table:spinconductivitysummary}).  Due to this high resistivity, Ho is unlikely to be useful for applications despite its relatively large spin-Hall torque ratio. 

The amplitude of the antisymmetric component of the ST-FMR signals gives additional information about the strength of the ``effective field'' component of the current-induced torque oriented perpendicular to the sample plane (see Eq.\ (2)). The Oersted field produced by the applied current contributes to this torque, but we also find a significant contribution from a spin-orbit torque that is oriented opposite to the Oersted field torque.  In the Ho samples, the spin-orbit contribution is actually larger than the Oersted field contribution, so that the sign of the antisymmetric ST-FMR component is reversed relative to the other samples [compare Fig.~\ref{fig:Ho_STFMR} with the other panels in Fig.~\ref{fig:STFMR}].  We can calculate the Oersted field based on our applied current and the measured resistivities of the sample layers, and then estimate from the amplitude of the antisymmetric ST-FMR component the out-of-plane spin torque ratio, with the results shown in Table~\ref{table:spinHallsummary}. (The negative signs indicate that the direction is opposite to the torque from the Oersted field.)  In all cases the magnitude of $\xi_{\perp,SO}$ is at least comparable to the antidamping spin-orbit torque ratio $\xi_{SH}$.    

\begin{table}[b] 
\caption{\label{table:spinconductivitysummary} Estimated lower bounds on the  spin torque conductivities ($\sigma^{exp}_{SH}$) based on this work, along with the measured electrical resistivity ($\rho_{RE}$) of each RE film.   A literature value for Pt measured in a Pt/Py bilayer system is included for comparison.  The measured value of resistivity for our Hf films is 72 $\mu\Omega$ cm, and for our Py films is 65 $\mu\Omega$ cm.}
\begin{tabular}{c c c c }
\hline
\hline
Element & t$_{Hf}$ (nm) & $\sigma^{\rm exp}_{SH}$ ($\hbar/2e$) 1/($\Omega$ cm)  & $\rho_{RE}$ ($\mu\Omega$ cm)  \\
\hline
Gd & 1.5 & 110 & 350 \\
Dy & 1.5 & 180 & 330 \\
Ho & 1.5 & 210 & 780 \\
Lu& 0 & 160 & 87 \\
\hline
Pt \cite{PhysRevLett.106.036601} & 0 & 3000 & 20\\
\hline
\hline
\end{tabular}
\end{table}

\begin{figure}[h!]
\includegraphics{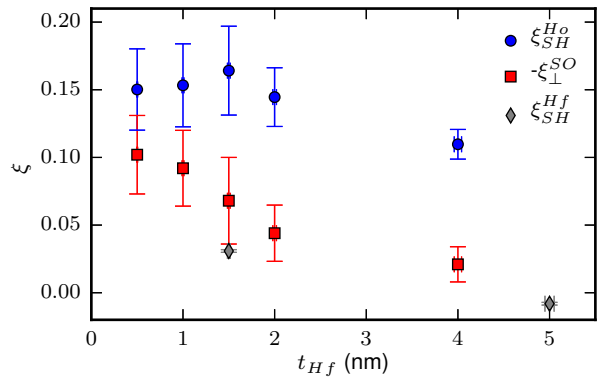}
\caption{(Color online) Experimental estimates of the in-plane spin Hall (blue circles) and out-of-plane effective field (red squares) torque ratios in the Ho (10 nm)/Hf (t$_{Hf}$ nm)/Py (5 nm)/AlO$_x$ multilayer structure using ST-FMR. For comparison, we also plot the in-plane torque ratios for measurements of a Hf (t$_{Hf}$ nm)/Py (5 nm)/AlO$_x$ multilayer for t$_{Hf}$= 1.5 nm and 5 nm.\label{fig:SHAvsthf}}
\end{figure}

For Ho, which provides the strongest spin-Hall torque from among the four elements, we fabricated a series of devices with Hf spacer thicknesses of 0.5, 1, 1.5, 2, and 4 nm, with the results for $\xi_{SH}$ and $\xi^{SO}_{\perp}$ shown in Fig.~\ref{fig:SHAvsthf}. Previous experiments from our research group have indicated that the spin Hall effect in Hf is very weak\cite{apl.hfenhances}, so we anticipated that both $\xi_{SH}$ and $\xi^{\perp,SO}$ would decay to small values as a function of increasing Hf thickness, over a characteristic scale determined by the spin diffusion length in Hf, $\approx$ 1-1.5 nm. (We note that other research groups have reported larger in-plane spin-torque ratios from Hf, presumably due to films with different crystal structures or different impurities \cite{apl/106/3/10.1063/1.4906352,apl/106/16/10.1063/1.4919108,apl/108/20/10.1063/1.4951674}, but our sputter chamber has always yielded negligible spin-torque ratios for Hf films.)  As a function of increasing Hf thickness we find that $\xi^{\perp,SO}$ behaves consistently with our expectation but, surprisingly, $\xi_{SH}$ does not. Instead, $\xi_{SH}$ decreases only slightly (by about 30\%) as the HF thickness is increased from 0.5 nm to 4 nm.  In contrast, control samples consisting of Hf($t_{Hf}$)/Py(5 nm)/AlO$_x$ with no RE layer and with $t_{Hf}$ = 1.5 nm and 5 nm did confirm a small value of $\xi_{SH}$ for Hf (Fig.~\ref{fig:SHAvsthf}), consistent with previous results from our group\cite{apl.hfenhances}.

We checked using x-ray diffraction whether there was any signficant structural change for the Hf within the Ho(10 nm)/Hf(4 nm)/Py(5 nm)/AlO$_x$ sample compared to the Hf(5 nm)/Py(5 nm)/AlO$_x$ control, and could observe no difference. We therefore suggest that the most likely explanation of of the unexpected behavior shown in Fig.~\ref{fig:SHAvsthf}, the relatively large values of $\xi_{SH}$ for thick Hf spacers in the Ho/Hf/Py/AlO$_x$ samples, is that the Ho and Hf layers may not be well-separated.  For example, a thin layer of Ho may migrate preferentially to the Py interface even for relatively thick Hf, or Ho impurities may exist within the Hf and contribute an extrinsic spin Hall effect.  Given the weak dependence of $\xi_{SH}$ as the Hf thickness is reduced from 1.5 nm to 0.5 nm (which should be well below the spin diffusion length), we still expect the the contribution from within the RE layer to dominate the spin Hall torque for the data in Fig.~II, even if there is some additional extrinsic spin Hall contribution.  Still, given this uncertainty with respect to the effect of the Hf spacer layer, we will not attempt in this initial survey to extrapolate our results to zero Hf thickness, but will simply state the measured spin-torque ratios for each of our samples. It is important to note that we are able to rule out any significant contribution to the spin-orbit torques from an interfacial effect at the Py/AlO$_x$ or the Hf/Py interfaces given the absence of any measurable spin-orbit torque in the Hf(5 nm)/Py(5 nm)/AlO$_x$ control sample.

The Ho samples with different Hf spacer thickness can also be used to analyze how the magnetic damping of the ferromagnetic layer is affected by the spacer. We discuss these data in an appendix.

\section{Theory calculations from first-principles}

\subsection{Theory}

The intrinsic spin Hall conductivity matrix elements, $\sigma^s_{\alpha,\beta}$, can be calculated  using the Kubo formula as:
\begin{align}
\sigma^s_{\alpha,\beta} =& -\frac{2e}{h}\int_{BZ} \frac{d^3k}{(2\pi)^3} \Omega^s_{\alpha,\beta}(k)
\label{eqn:spin_hall_conductivity}\\
\Omega^s_{\alpha,\beta}(k) =& -2\text{Im} \left[\sum_n f_n(k) \sum_{m\neq n}\frac{j^s_{\alpha,nm}v_{\beta,mn}}{(\omega_m - \omega_n)^2}\right]\\
\hat{j}^s_\alpha =& \frac{1}{2} \{\hat{s}_s,\hat{v}_\alpha\},
\end{align}
where $\hat{j}^s$, $\hat{s}$ and $\hat{v}$ are the spin current, spin, and velocity operators, respectively. We calculate these quantities for Gd, Dy, Ho, and Lu using first-principles, density-functional theory (DFT) in both the local density approximation (LDA) and the local density plus Hubbard U approximation (LDA+U) to the exchange-correlation fuctional. We considered only the intrinsic contribution to spin Hall conductivity, thereby ignoring possible contributions from extrinsic impurity scattering. 

Because of the hexagonal close packed (HCP) crystal structure of the the rare-earths studied, the intrinsic spin Hall conductivity is predicted to be anisotropic. Since the samples used in experiment are polycrystalline with random orientation, we expect that the experimentally measured spin Hall conductivities are averaged over an isotropic distribution of all orientations of the crystal. To capture this theoretically, we  present the average of $\sigma^z_{x,y}$, $\sigma^x_{y,z}$ and $\sigma^y_{z,x}$, which we call $\sigma^{\rm DFT}_s$. Furthermore, we note that the quantity calculated theoretically is not identical to  the spin Hall torque conductivity $\sigma^{\rm exp}_{SH}$ measured experimentally. As mentioned above, the experimentally determined value for $\sigma^{\rm exp}_{SH}$ is a lower bound as it will be reduced by any less-than ideal spin transmission through the sample interfaces and the Hf spacer layer in the real samples. The calculated spin Hall conductivity $\sigma^{\rm DFT}_{SH}$ does not include any details about transmission of spin current into a ferromagnet and represents the case of perfect spin transmission.

\subsection{DFT details and approach}

Rare-earth metals pose a significant challenge for first-principles theories due to the degree of localization of the 4$f$ electrons. To investigate this issue, we constructed two different sets of norm conserving, fully-relativistic pseudopotentials (PsP's) in the local density approximation (LDA), following Perdew and Wang \cite{PhysRevB.45.13244,PhysRevB.23.5048}. These pseudopotentials were constructed using the Atomic Pseudopotential Engine (APE) \cite{Oliveira2008524}. The pseudopotentials generated were benchmarked against the fully-relativistic all-electron potential.

One set of PsP's posits that 4$f$ electrons are highly localized, and as such do not affect the electronic properties of rare-earths \cite{JMMM.61.139,JLCM.158.207}. In this case, the 4$f$ states are frozen in the core of the PsP and the valence only contains the 6$s$, 5$d$ and 5$p$ electrons. We refer to calculations using these PsP's, which were performed within the LDA, as `$f$-in-core'. The second set of PsP's are constructed with the 4$f$ states in the valence such that they can in principle affect the electronic properties \cite{Nat.399.6738}. We refer to calculations using these PsP's as `$f$-in-band'. These were performed within the LDA+U. Since the appropriate values of the Hubbard U$_f$ parameter are uncertain \cite{Topsakal2014263,PhysRevB.37.10674}, we perform such calculations for a range of $U_f$ values.

We performed the calculations using Quantum ESPRESSO \cite{QE-2009}. Convergence of total energy of our structures required a plane-wave energy cut off of 50 Ry for the `$f$-in-core' PsP's and 290 Ry for the `$f$-in-band' PsP's. A $k$ mesh of 15x15x10 was used for the self-consistent part of our DFT calculations to obtain electronic ground states. The crystal structures for Gd, Dy, Ho and Lu are in all cases hexagonal close-packed (hcp) \cite{Mohanta20101789}. We fully relax all structures.  

Calculating the spin Hall conductivity accurately requires a very dense $k$ mesh. To circumvent this issue we mapped our DFT ground state wavefunctions onto a maximally localized Wannier Function basis using WANNIER90 \cite{Mostofi2008685}. We  extracted the relevant matrices for calculation of spin Hall conductivity, including the velocity matrix $\hat{v}$ and spin matrix $\hat{s}$ of the system. Finally, we employed an adaptive $k$ mesh strategy, inspired by the anomalous Hall functionality in WANNIER90 \cite{PhysRevB.74.195118}, to calculate the spin Hall conductivity matrix.

\subsection{Spin Hall conductivity in the `f-in-core' picture}

\begin{figure}[h!]
	\subfloat{
        \includegraphics{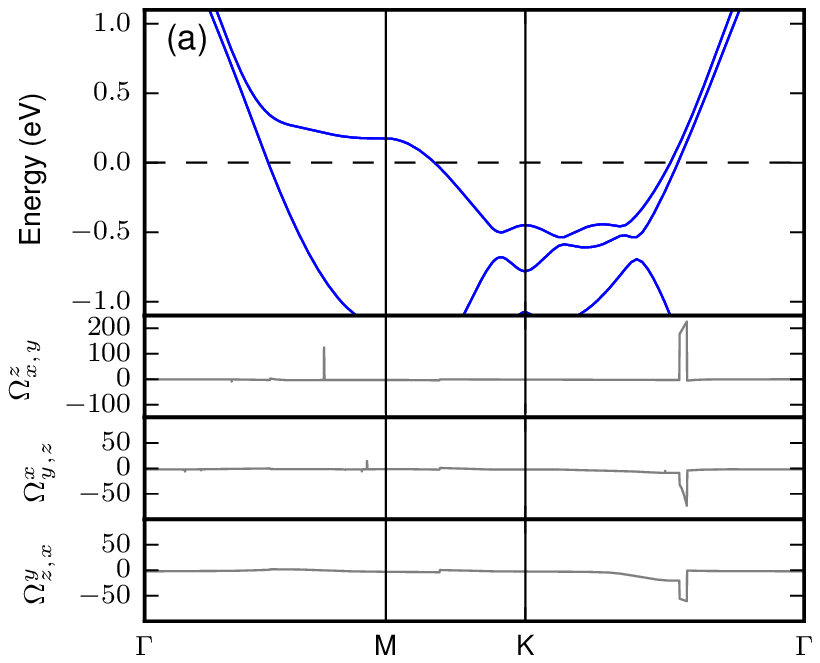}
        \label{fig:Lusdband}
        }  
        
    \subfloat{
        \includegraphics{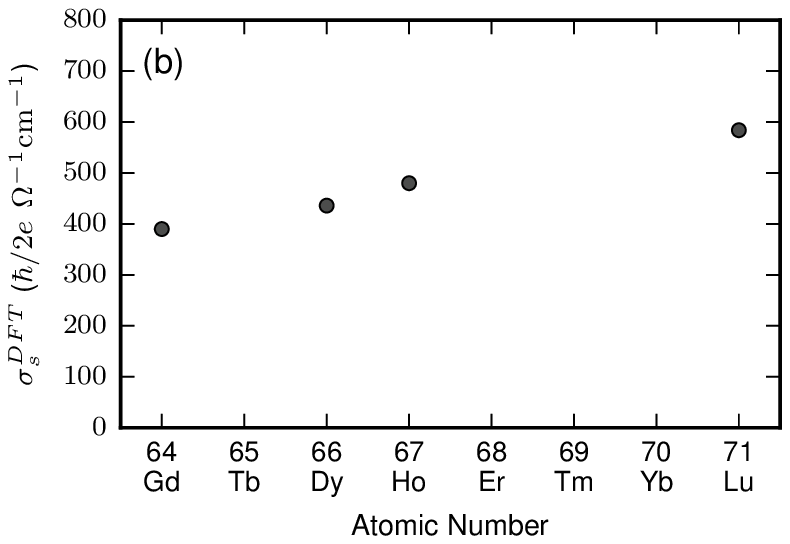}
        \label{fig:SHCsd}
        }
\caption{(Color online) For the `f-in-core' picture: (a) Band structure for Lu (with Fermi energy $\epsilon_f$ at 0 eV) and spin Berry curvature $\Omega^s_{\alpha,\beta}(k)$ (b) calculated spin Hall conductivities. \label{fig:Lusd}}
\end{figure}

\begin{figure*}[t!]
\includegraphics{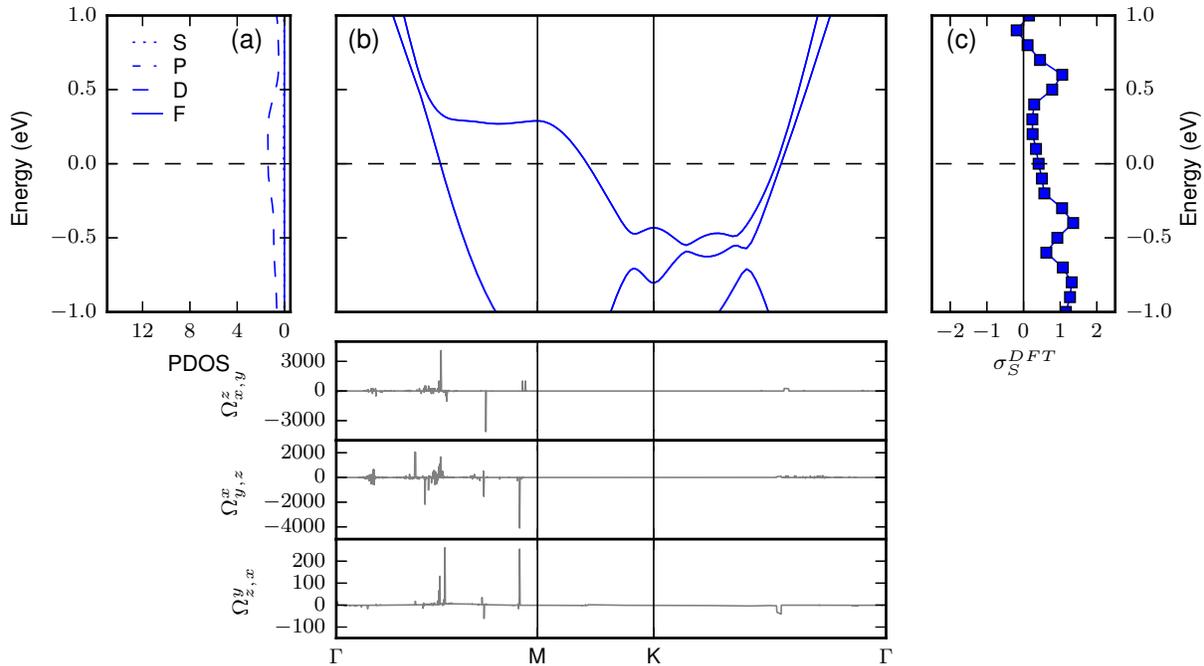}
\caption{(Color online) For Lu in the `f-in-band' picture: (a) Projected density of states (`PDOS' (states/eV-atom)) (b) the band structure (top) and spin Berry curvature $\Omega^s_{\alpha,\beta}(k)$ (bottom) with U$_{\mathrm{f}}$=5.5 eV and Fermi energy $\epsilon_f$ at 0 eV (c) calculated spin Hall conductivity ($\sigma^{DFT}_{S}$ ($\hbar/2e$ $\,\mathrm{m}\Omega^{-1}$$\mathrm{cm}^{-1}$)) vs. Fermi level.}\label{fig:Lu-s-d-f}
\end{figure*}

Results of the spin Hall conductivity calculations within the `f-in-core' picture for Gd, Dy, Ho, and Lu are shown in Fig.~\ref{fig:SHCsd}, which shows a monotonic increase in $\sigma^{\rm DFT}_s$ with atomic number $Z$. This trend can be understood by first noting that all four elements have the same crystal structure (hcp) and, within the `f-in-core' picture, have the same electron filling, i.e., $6s^2 5d^1$, resulting in a very similar electronic structure. Therefore, the monotonic increase in spin Hall conductivity from Gd to Lu is consistent with an increase in the strength of spin-orbit coupling, a result shown previously by Ref.~\onlinecite{doi:10.1143/JPSJ.76.103702} for Pt.

To understand the origin of the spin Hall conductivity in the `f-in-core' picture we focus on the electronic structure of Lu, shown in  Fig.~\ref{fig:Lusdband}, as representative. We find that there are $5d,6s$ bands near the Fermi level $\epsilon_F$, with avoided band crossings between the $\Gamma$ and K points. These avoided crossings are the primary source of the spin Berry curvature $\Omega^s_{\alpha,\beta}(k)$, also plotted in Fig.~\ref{fig:Lusdband}, which upon integration over all occupied $k$-states gives the spin Hall conductivity $\sigma^s_{\alpha,\beta}$. These features of the electronic structure do not vary for the different elements within the `f-in-core' picture.

\subsection{Spin Hall conductivity in the `f-in-band' picture}

\begin{figure*}[t!]
\includegraphics{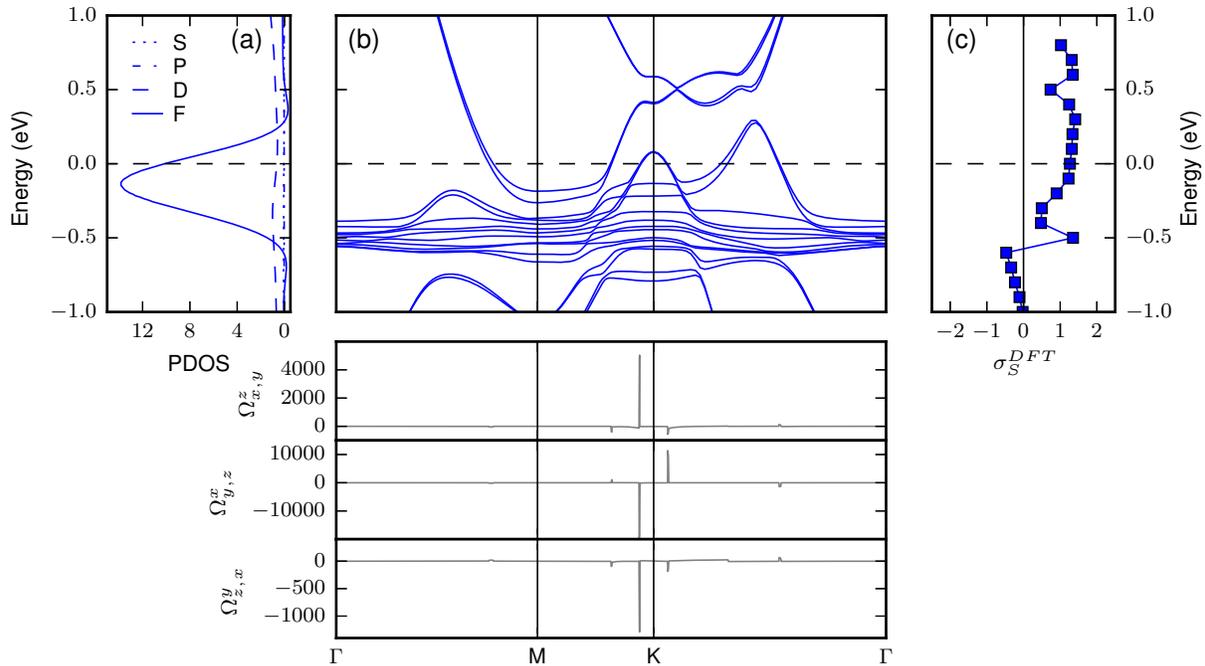}
\caption{(Color online) For Ho in the `f-in-band' picture: (a) Projected density of states (`PDOS' (states/eV-atom)) (b) the band structure (top) and spin Berry curvature $\Omega^s_{\alpha,\beta}(k)$ (bottom) with U$_{\mathrm{f}}$=4.9 eV and Fermi energy $\epsilon_f$ at 0 eV (c) calculated spin Hall conductivity ($\sigma^{DFT}_{S}$ ($\hbar/2e$ $\,\mathrm{m}\Omega^{-1}$$\mathrm{cm}^{-1}$)) vs. Fermi level.}\label{fig:Ho-s-d-f}
\end{figure*}

For `f-in-band' calculations, we included the $4f$ electrons in the variational density function and applied a Hubbard potential $U_f$ to account for correlation effects. The values of the Hubbard parameter for Gd, Dy, Ho and Lu were taken to be 4.6, 5.0, 4.9, and 5.5 respectively \cite{Topsakal2014263}. (We discuss the consequences of assuming different values of $U_f$ below.) To highlight the effect of $4f$ electrons on the electronic structure we will compare and contrast Lu and Ho. This is because, from among the four elements we examined, Lu with its full $4f$-shell is expected to display the least contribution of $4f$ electrons to the band at the Fermi level, while Ho is expected to display the most. Dy has a larger $U_f$ than Ho, and Gd has a more stable $f^7$ shell filling, reducing the expected presence of $4f$ bands near $\epsilon_F$ as compared to Ho.

    We first focus on Lu, for which the projected density of states and electronic structure are shown in Figs.~\ref{fig:Lusd}a and ~\ref{fig:Lusd}b, respectively . The electronic structure of Lu in this `f-in-band' picture (Fig.~\ref{fig:Lu-s-d-f}b) is very similar to the one obtained in the `f-in-core' picture (Fig.~\ref{fig:Lusd}a). We restrict the range of the `f-in-band' figures to 1 eV above and below the Fermi level $\epsilon_F$ to limit the computational complexity. The complete and stable $f^{14}$ filling of the $4f$-level in Lu and the large Hubbard term ($U_f$ = 5.5 eV) help place the $4f$ bands deep below  $\epsilon_F$, \textit{i.e.} $\sim$ 5 eV below. This leaves the bands near the Fermi level to originate from $5d, 6s$ electrons alone and makes them very similar to the bands obtained in the earlier `f-in-core' calculations. As a result, the spin Hall conductivity of Lu from the `f-in-band' calculations for $U_f = 5.5$ eV, 410 ($\hbar/2e$) $\Omega^{-1}$ cm$^{-1}$, is similar to the result  580 ($\hbar/2e$) $\Omega^{-1}$ cm$^{-1}$ from the `f-in-core' calculation. The difference between these values arises from the difference in the two pseudopotentials.

    The influence of the $f$ electrons in Ho is very different than in Lu. The projected density of states and electronic structure for Ho are shown in Figs.~\ref{fig:Ho-s-d-f}a and ~\ref{fig:Ho-s-d-f}b, respectively. While we can identify the signatures of bands originating in $6s$ and $5d$ electrons, similar to those calculated in `f-in-core' calculations, these bands are now pushed about 1.0 eV higher in energy. In addition, a host of flat bands originating from the $4f$ levels are present centered near -0.5 eV, giving rise to a large peak in the density of states (DOS)  arising from $4f$-electrons ranging in energy from -0.6 to 0.3 eV, with a significant presence at the Fermi level. The largest contribution to spin Berry curvature, shown in Fig.~\ref{fig:Ho-s-d-f}b, comes from the avoided band crossings around the K point. The calculated spin Hall conductivity of Ho, in the `f-in-band' framework is 1260 ($\hbar/2e$) $\Omega^{-1}$ cm$^{-1}$, which is almost double the value obtained in the `f-in-core' calculation, 640 ($\hbar/2e$) $\Omega^{-1}$ cm$^{-1}$. Such a large difference between $\sigma_{s,f\mathrm{-in-core}}^{DFT}$ and $\sigma_{s, f\mathrm{-in-band}}^{DFT}$ cannot be explained by a difference in the pseudopotentials alone. The central result of this calculation is therefore that, within our theoretical framework, $4f$-electrons do contribute to the spin Hall conductivity of Ho in a significant way. 
    
    Differences in $\sigma_s^{DFT}$ between the `f-in-band' and `f-in-core' calculations can arise, in principle, not only from a direct contribution of the $f$-electron bands, but also from $5d$-$4f$ interactions that will modify the band structure.  Since the $4f$ electrons are largely localized and the $5d,6s$ bandstructure is altered only to a small extent by the presence of $4f$-electrons, we conclude that these interactions are small. This suggests that most of the difference in $\sigma_s^{DFT}$ arises from the $4f$ electron contribution, meaning that this $4f$ contribution  is large and positive for Ho. This theoretical result agrees with our intuitive discussion of the spin Hall effect from `LS' coupling detailed earlier in Fig.~\ref{fig:LS} that in the late lanthanides the $4f$ bands should make a positive contribution to the total spin Hall conductivity.
   
    Our DFT calculations for Dy and Gd also show the signatures of shifted $5d,6s$ band-structure features and flat $4f$ bands. In the `f-in-band' picture, the f-DOS at the Fermi level increases in the order Gd, Dy, and Ho. Additionally, the atomic number Z and thus the spin orbit coupling constant also increases in the same order. This results in an increasing trend of spin Hall conductivity from Gd to Dy to Ho. Despite the larger atomic number Z and spin orbit coupling constant displayed by Lu as compared to Ho, the complete absence of $4f$ DOS at $\epsilon_f$ leads to a significantly smaller spin Hall conductivity for Lu compared to Ho. Altogether, the spin Hall conductivity trend predicted by the `f-in-band' calculations is  $\sigma_{s,Gd}^{DFT} < \sigma_{s,Dy}^{DFT} < \sigma_{s,Ho}^{DFT} > \sigma_{s,Lu}^{DFT}$, which is in contrast with the prediction of the `f-in-core' framework for a monotonic increase with atomic number Z. The former agrees with the trend we observe in our experiments.

\begin{figure}
\includegraphics{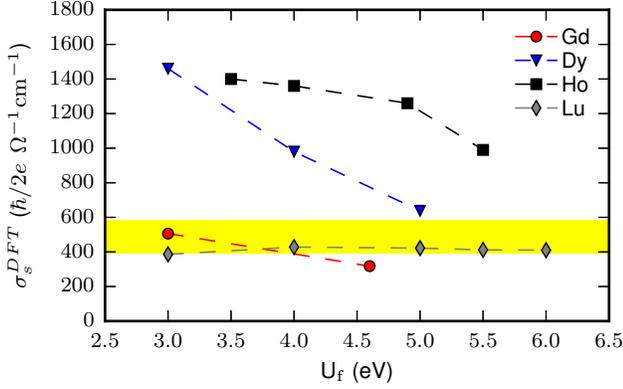}
\caption{(Color online) The calculated spin Hall conductivity, $\sigma^{\rm DFT}_{s}$, for Dy, Ho and Lu using DFT-LDA with varying U$_{\mathrm{f}}$ (`$f$-in-band'). As the $4f$ states become more localized with increasing U$_{\mathrm{f}}$, we find that the calculated $\sigma^{DFT}_{s}$ decreases for both Dy and Ho. For Gd and Lu we find only a modest change in the calculated $\sigma^{DFT}_{s}$.
The yellow region spans the values of $\sigma^{DFT}_{s}$ calculated in the  `$f$-in-core' picture (Fig.~\ref{fig:SHCsd})}\label{fig:SHCvsU}
\end{figure}

Given that there is some uncertainty in the literature about the most appropriate values to use for $U_f$, we also investigated how changes in $U_f$ affect the calculated spin Hall conductivity, shown in Fig.~\ref{fig:SHCvsU}. We find that generally $\sigma_s^{DFT}$ decreases with increasing U$_f$. In light of the above discussion, this reflects that a higher U$_{f}$ will promote greater localization, pushing the $4f$ levels well below $\epsilon_F$ and lessening their participation in the conduction bands. This diminishes any enhancement to $\sigma_s^{DFT}$ coming from the $4f$ levels so that in the limit of large U$_{f}$ the `$f$-in-band' $\sigma_s^{DFT}$ reduces approximately to the `$f$-in-core' $\sigma_s^{DFT}$ values (the region of the yellow band in Fig.~\ref{fig:SHCvsU}). We  find that in the cases of Dy and Ho, $\sigma_s^{DFT}$ depends quite sensitively upon the choice of U$_{f}$. For Dy in particular, decreasing $U_f$ from 5 eV to 3 eV doubles the calculated spin Hall conductivity. This sensitivity to U$_f$, which depends on the details of electronic structure, highlights how potently the $4f$ bands can enhance $\sigma_s^{DFT}$. For our main results, we take values of Hubbard $U_f$ for Gd, Dy, Ho and Lu to be 4.6, 5.0, 4.9, and 5.5 eV from the work of Topsakal and Wentzcovitch \cite{Topsakal2014263}. In contrast, van der Marel and Sawatzky \cite{PhysRevB.37.10674} propose that $U_f$ for Gd, Dy, Ho, and Lu should be 11, 5.8, 6.8, and 7 eV, respectively. We account for these differences by including estimated error bars in the plot of our final theoretical predictions for $\sigma_s^{DFT}$, discussed below and in Fig.~\ref{fig:summaryfigure}. 
   
\begin{figure}
\includegraphics{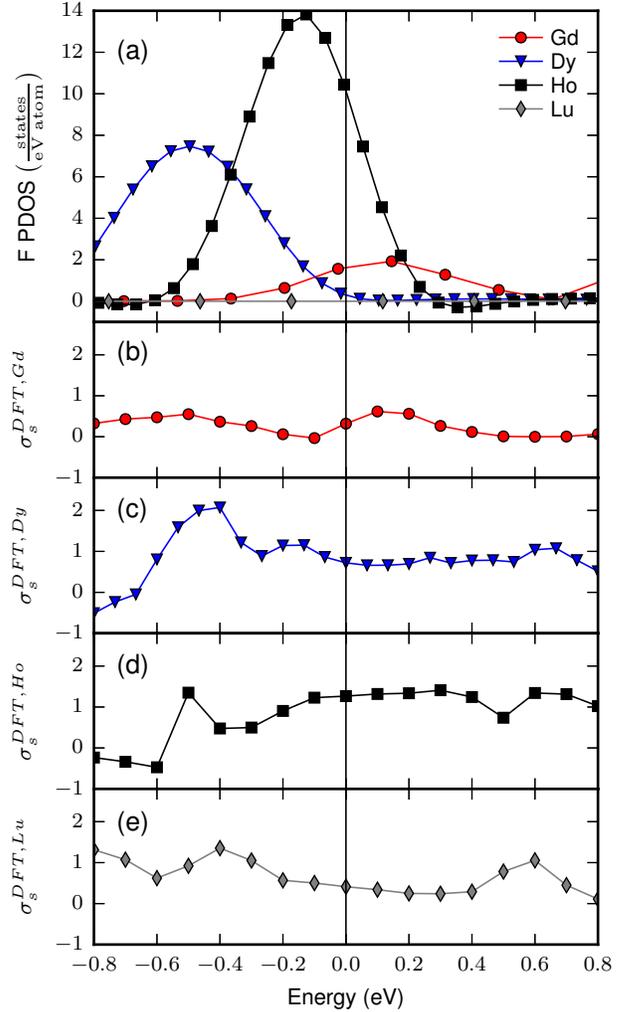}
\caption{(Color online) (a) Projected density of states of Gd, Dy, Ho, and Lu. (b)-(e) $\sigma_s^{DFT}$ ($\hbar/2e$ m$\Omega^{-1}$ cm$^{-1}$) vs Fermi level $\epsilon_f$ for each element.}\label{fig:pdosfermiscan}
\end{figure}
   
    We can further analyze the $4f$-electron contributions to spin Hall conductivity by comparing the energy dependence of the DOS for the $f$ electrons (f-DOS) with the variation of spin Hall conductivity as a function of Fermi level $\epsilon_F$, as plotted in Fig.~\ref{fig:pdosfermiscan}. A correlation between spin Hall conductivity and the f-DOS is evident in the plot for Dy. Around -0.5 eV, both f-DOS and spin Hall conductivity for Dy attain their maximum value with the latter reaching as high as 2070 ($\hbar/2e$) $\Omega^{-1}$ cm$^{-1}$. This is almost 3 times the value observed at the native Fermi level. The correlation of spin Hall conductivity with the f-DOS is less clear for Ho. This is likely due to the proximity of $4f$ to $5d$ bands and the subsequent hybridization between them. The largest value of spin Hall conductivity for Ho in the energy range we have examined is 1350 ($\hbar/2e$) $\Omega^{-1}$ cm$^{-1}$, which is close to the value for the native Fermi level. For Lu, the $4f$ bands lie outside of our scanning region of 0.8 eV around the Fermi level and the variation in spin Hall conductivity in our plotting range arises from the $5d$-electrons alone. The spin Hall conductivity peak for Lu at -0.4 eV and for Dy and Ho near 0.6 eV originates from a $5d$ band feature that is shifted to different energies in the two elements because of different interactions with the f levels. In the case of Gd, there is a small but finite presence of f-DOS at the Fermi level. This coincides with the presence of the $5d$ band feature which is responsible for the spin Hall conductivity peak at -0.4 eV in Lu. Together the two contribute to the small spin Hall conductivity value predicted by our calculations.
    This analysis suggests that engineering the placement of the Fermi level, possibly by alloying, could enhance the spin Hall conductivity significantly if $\epsilon_F$ were tuned to place it in better proximity to either a peak in the $4f$ DOS or to an avoided crossing of the $5d,6s$ bands. The largest value of $\sigma_s^{DFT}$ we have found by tuning $\epsilon_F$ in our four elements is 2070 ($\hbar/2e$) $\Omega^{-1}$ cm$^{-1}$ ($\approx\frac{2}{3}$ that of Pt \cite{PhysRevLett.106.036601}),  for Dy at $\epsilon_F = -0.5$ eV. 
  
\section{Comparison between Experiments and Theory}

\begin{figure}
\includegraphics{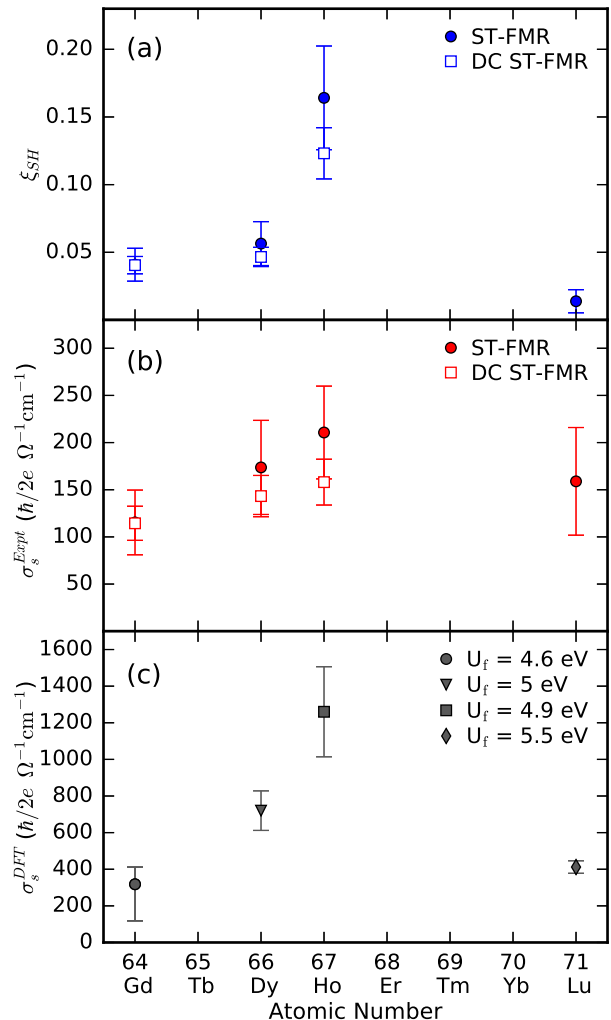}
\caption{(Color online) Experimentally measured (a) spin Hall torque ratio and (b) spin Hall torque conductivity of Gd, Dy, Ho, and Lu. The Gd, Dy, and Ho devices have a 1.5 nm layer of Hf between the RE metal and ferromagnet. (c) Spin Hall conductivities for the same elements calculated using DFT+U with values of U$_{\mathrm{f}}$ taken from the work of Topsakal and Wentzcovitch \cite{Topsakal2014263}.\label{fig:summaryfigure}}
\end{figure}

We compare in Fig.~\ref{fig:summaryfigure} our experimental results of spin torque ratio and spin torque per unit electric field $\sigma^{\rm exp}_{SH}$ with our theoretical predictions for spin Hall conductivity $\sigma^{\rm DFT}_s$.  For best consistency in the experimental values, we have used the results for Gd, Dy, and Ho with a 1.5 nm Hf spacer.  The Lu sample had no Hf spacer. 

As discussed above and shown in Fig.~\ref{fig:SHCsd}, if there were no $f$ orbital participation in the spin Hall effect, we should expect to see only a small monotonic increase in the spin Hall conductivity from Gd to Lu owing to the increasing spin orbit coupling with Z. Instead, in the experiment we observe a quite sizable increase in measured spin Hall effect going from Gd to Dy to Ho, and then a drop to a much smaller value in Lu that cannot be captured by the `f-in-core' picture. In contrast, the `f-in-band' calculations are in good qualitative agreement with this tell-tale drop. The `f-in-band' calculations can explain the strong variations in the spin Hall conductivity as due in large part to the differing magnitudes of the $4f$ density of states near the Fermi level for the different elements.

Quantitatively, the `f-in-band' calculations predict values of $\sigma_{s}^{DFT}$ that are significantly larger than the experimentally-measured quantities $\sigma_{SH}^{exp}$, by a factor of 3-6 depending on the element. That $\sigma_{s}^{DFT}$ is larger than $\sigma_{SH}^{exp}$ is not surprising, because the experimentally-measured torques will be reduced by less-than-ideal spin transmission through sample interfaces and by the presence of the Hf spacer in the Gd, Dy, and Ho samples. The large value of the difference suggests that spin scattering at the rare-earth interfaces may cause significant attenuation of the spin currents generated by the spin Hall effect.  Another indication of a large amount of spin scattering at interfaces is the fact that the magnetic damping in the RE/Py and RE/Hf/Py samples is generally much larger than can be explained by the standard theory of spin pumping from the Py layer (see the Appendix).  Improvement of the rare-earth interfaces might therefore be able to reduce this spin scattering and enable stronger spin-orbit torques. 

We conclude, based on both the measurements and the `f-in-band' calculations, that the $4f$ electrons do indeed contribute to an enhancement of the spin Hall effect in Ho, and perhaps to a lesser extent in Dy.  The amount of enhancement depends on the proximity of the $4f$-electrons to the Fermi level, so that the spin Hall effect in Gd and Lu remains small. We suggest that the spin Hall conductivity might be further enhanced by tuning the Fermi level closer to either the peak in the $4f$ DOS or to a $5d$ avoided band crossing present in the lanthanides. Another mechanism of enhancing SHE in rare-earths could be to promote the delocalization of the $4f$ electrons.

\section*{Acknowledgments}
We thank Gregory Stiehl, David MacNeill, Di Xiao, and Craig Fennie for helpful discussions.

This work was supported by the National Science Foundation (DMR-1406333 and through the Cornell Center for Materials Research, part of the NSF MRSEC program, DMR-1120296) and the Office of Naval Research.  We made use of the shared facilities of the Cornell Center for Materials Research and the Cornell Nanoscale Facility, which is supported by the NSF (ECCS-1542081). 

\section{Appendix}
\subsection{Analysis of Magnetic Damping}

 As noted in the main text, the magnetic damping is increased for heterostructures in which a RE layer is placed in proximity to the ferromagnetic layer, even if there is a thin Hf spacer in between.  This increase reflects the contributions to magnetic dissipation from spin pumping into the RE layer and any spin relaxation that may occur at the Hf/RE interface.  For Gd, Dy, and Ho, we compared the value of the Gilbert damping ($\alpha_G$) determined by ST-FMR for RE (10 nm)/Hf (1.5 nm)/Py (5 nm)/AlO$_x$ multilayers with the value ($\alpha_G-\Delta \alpha$) for control samples without the RE layer, but still containing the Hf spacer layer.  Assuming diffusive spin flow, we can define an effective spin mixing conductance for spin transport from the Py/Hf to the RE as \cite{PhysRevB.92.064426} 
\begin{equation}
g_{eff}^{'\uparrow\downarrow}=\frac{M_st_{FM}^{eff}}{\gamma\hbar}\Delta \alpha. \label{eqn:geff}
\end{equation}
The results are listed in Table~\ref{table:spinmixing}. The values determined by this technique should be less than the true effective mixing conductance of the interface between Py and Hf because of the spin current attenuation within the Hf.  In spite of this, the values for Dy and Ho (in particular) are surprisingly large. In the absence of interfacial spin relaxation mechanisms, the maximum effective spin mixing conductance should be no more than $2 \rho_{RE}/\lambda_{RE}$, where $\rho_{RE}$ is the electrical resistivity of the rare-earth material and $\lambda_{RE}$ is its spin diffusion length, because this quantity determines the ability of the RE layer to absorb a spin current from the precessing ferromagnet. If we assume $\lambda_{RE} \approx 1$ nm, we find for Dy and Ho that $g_{eff}^{'\uparrow\downarrow}$ is much larger than $2 \rho_{RE}/\lambda_{RE}$ (see Table~\ref{table:spinmixing}). We therefore conclude that the large values of damping in the Dy and Ho samples are not dominated by conventional spin pumping from the ferromagnet into the RE layer, but rather are more likely the result of interfacial intermixing, and the ability of Dy and Ho impurities to greatly increase spin relaxation \cite{IEEEtransmagn.37.2001,apl/82/8/10.1063/1.1544642}.

\begin{table}
\caption{\label{table:spinmixing} A summary of the damping enhancement $\Delta\alpha$, the calculated effective spin mixing conductance for each RE (10 nm)/Hf (1.5 nm)/Py (5 nm)/AlO$_x$ heterostructure as described in the Appendix, and an estimated effective spin conductivity $2 \rho_{RE}/ \lambda_{RE}$ for each rare-earth. $\alpha_0$ = 0.0077 as measured for a sapphire/Hf (1.5 nm)/Py (5 nm)/AlO$_x$ control sample. $\lambda_{RE}$ is assumed to be $\approx$ 1 nm.}
\begin{tabular}{c c c c }
\hline
\hline
Element  & $\Delta\alpha$ ($10^{-3}$)  & $g_{eff}^{'\uparrow\downarrow}$ (nm$^{-2}$) & $2 \rho_{RE}/\lambda_{RE}\frac{e^2}{h}$ (nm$^{-2}$)\\
\hline
Gd 	& 1.2	 &	2.9 	&	9.6\\
Dy 	& 9.3 	 &	18.6	& 	10.6\\
Ho 	& 2.4	 &	6.2 	&	4.4\\
\hline
\hline
\end{tabular}
\end{table}

\begin{figure}[b]
\includegraphics{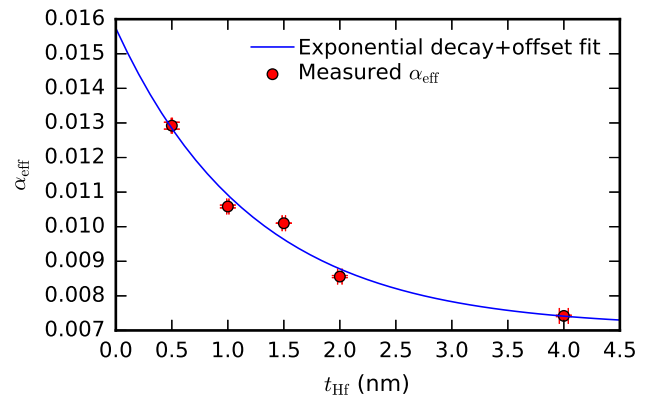}
\caption{(Color online) Damping of Ho (10 nm)/Hf ($t_{Hf}$)/Py (5 nm)/AlO$_x$ stacks for  $t_{Hf}$ = 0.5, 1, 1.5, 2, and 4 nm (red circles). Fitting to $\alpha=$offset+A$e^{-t/\lambda_{sd}}$ (blue line) yields offset = 0.0071, A = 0.0087, and $\lambda_{sd}$=0.81 nm.}\label{fig:Hf_thicknesses}
\end{figure}

We also measured the effective Gilbert damping coefficient as a function of Hf spacer thickness in a set of Ho (10 nm)/Hf ($t_{Hf}$)/Py (5 nm)/AlO$_X$ samples for  $t_{Hf}$ = 0.5, 1, 2, and 4 nm (Fig.~\ref{fig:Hf_thicknesses}).  The results fit well to a simple exponential function plus an offset, with an effective spin diffusion length in the Hf of 0.81 nm. The offset of 0.0071 gives the limit in which the Hf is so thick that the Ho layer is not relevant to the damping enhancement. This value agrees with the damping for the control sample of sapphire/Hf (1.5 nm)/Py (5 nm)/AlO$_X$ of $\alpha_0$ = 0.0077 that was used to determine the values of $\Delta\alpha$ in Table~\ref{table:spinmixing}. 

\end{document}